\documentclass[dvips]{article}
\usepackage{supertabular,lscape,epsfig}
\usepackage{amssymb}
\usepackage{amsmath}

\DeclareSymbolFont{ppa}{OT1}{ppl}{m}{it}
\DeclareMathSymbol{\vv}{\mathalpha}{ppa}{'166}

\begin{document}

\newcommand{\dd}{\,{\rm d}}
\newcommand{\ie}{{\it i.e.},\,}
\newcommand{\etal}{{\it et al.\ }}
\newcommand{\eg}{{\it e.g.},\,}
\newcommand{\cf}{{\it cf.\ }}
\newcommand{\vs}{{\it vs.\ }}
\newcommand{\zdot}{\makebox[0pt][l]{.}}
\newcommand{\up}[1]{\ifmmode^{\rm #1}\else$^{\rm #1}$\fi}
\newcommand{\dn}[1]{\ifmmode_{\rm #1}\else$_{\rm #1}$\fi}
\newcommand{\upd}{\up{d}}
\newcommand{\uph}{\up{h}}
\newcommand{\upm}{\up{m}}
\newcommand{\ups}{\up{s}}
\newcommand{\arcd}{\ifmmode^{\circ}\else$^{\circ}$\fi}
\newcommand{\arcm}{\ifmmode{'}\else$'$\fi}
\newcommand{\arcs}{\ifmmode{''}\else$''$\fi}
\newcommand{\MS}{{\rm M}\ifmmode_{\odot}\else$_{\odot}$\fi}
\newcommand{\RS}{{\rm R}\ifmmode_{\odot}\else$_{\odot}$\fi}
\newcommand{\LS}{{\rm L}\ifmmode_{\odot}\else$_{\odot}$\fi}

\newcommand{\Abstract}[2]{{\footnotesize\begin{center}ABSTRACT\end{center}
\vspace{1mm}\par#1\par
\noindent
{~}{\it #2}}}

\newcommand{\TabCap}[2]{\begin{center}\parbox[t]{#1}{\begin{center}
  \small {\spaceskip 2pt plus 1pt minus 1pt T a b l e}
  \refstepcounter{table}\thetable \\[2mm]
  \footnotesize #2 \end{center}}\end{center}}

\newcommand{\TableSep}[2]{\begin{table}[p]\vspace{#1}
\TabCap{#2}\end{table}}

\newcommand{\FigCap}[1]{\footnotesize\par\noindent Fig.\  %
  \refstepcounter{figure}\thefigure. #1\par}

\newcommand{\TableFont}{\footnotesize}
\newcommand{\TableFontIt}{\ttit}
\newcommand{\SetTableFont}[1]{\renewcommand{\TableFont}{#1}}

\newcommand{\MakeTable}[4]{\begin{table}[htb]\TabCap{#2}{#3}
  \begin{center} \TableFont \begin{tabular}{#1} #4 
  \end{tabular}\end{center}\end{table}}

\newcommand{\MakeTableSep}[4]{\begin{table}[p]\TabCap{#2}{#3}
  \begin{center} \TableFont \begin{tabular}{#1} #4 
  \end{tabular}\end{center}\end{table}}

\newenvironment{references}%
{
\footnotesize \frenchspacing
\renewcommand{\thesection}{}
\renewcommand{\in}{{\rm in }}
\renewcommand{\AA}{Astron.\ Astrophys.}
\newcommand{\AAS}{Astron.~Astrophys.~Suppl.~Ser.}
\newcommand{\ApJ}{Astrophys.\ J.}
\newcommand{\ApJS}{Astrophys.\ J.~Suppl.~Ser.}
\newcommand{\ApJL}{Astrophys.\ J.~Letters}
\newcommand{\AJ}{Astron.\ J.}
\newcommand{\IBVS}{IBVS}
\newcommand{\PASP}{P.A.S.P.}
\newcommand{\Acta}{Acta Astron.}
\newcommand{\MNRAS}{MNRAS}
\renewcommand{\and}{{\rm and }}
\section{{\rm REFERENCES}}
\sloppy \hyphenpenalty10000
\begin{list}{}{\leftmargin1cm\listparindent-1cm
\itemindent\listparindent\parsep0pt\itemsep0pt}}%
{\end{list}\vspace{2mm}}

\def\TYLDA{~}
\newlength{\DW}
\settowidth{\DW}{0}
\newcommand{\dw}{\hspace{\DW}}

\newcommand{\refitem}[5]{\item[]{#1} #2%
\def\REFARG{#3}\ifx\REFARG\TYLDA\else, {\it#3}\fi
\def\REFARG{#4}\ifx\REFARG\TYLDA\else, {\bf#4}\fi
\def\REFARG{#5}\ifx\REFARG\TYLDA\else, {#5}\fi.}

\newcommand{\Section}[1]{\section{#1}}
\newcommand{\Subsection}[1]{\subsection{#1}}
\newcommand{\Acknow}[1]{\par\vspace{5mm}{\bf Acknowledgements.} #1}
\pagestyle{myheadings}

\newfont{\bb}{ptmbi8t at 12pt}
\newcommand{\xrule}{\rule{0pt}{2.5ex}}
\newcommand{\xxrule}{\rule[-1.8ex]{0pt}{4.5ex}}
\def\thefootnote{\fnsymbol{footnote}}
\begin{center}
{\Large\bf The Optical Gravitational Lensing Experiment.\\
\vskip3pt
Stellar Distance Indicators in the Magellanic Clouds and Constraints on
the Magellanic Cloud Distance Scale\footnote{Based on observations obtained
with the 1.3~m Warsaw telescope at the Las Campanas Observatory of the
Carnegie Institution of Washington.}}
\vskip1cm
A.~~U~d~a~l~s~k~i
\vskip3mm
Warsaw University Observatory, Al.~Ujazdowskie~4, 00-478~Warszawa, Poland\\
e-mail: udalski@astrouw.edu.pl
\end{center}

\Abstract{{\it BVI} photometry of the Magellanic Clouds collected during the 
OGLE-II microlensing experiment makes it possible to study in detail 
photometric properties of the "major" stellar distance indicators in the 
Magellanic Clouds. In addition to Cepheids, RR~Lyr and red clump stars, which 
photometry was presented in the earlier OGLE papers, we present the so far 
most accurate determination of brightness of the tip of the red giant 
branch in the LMC and SMC. 

We analyze the ratios of brightness of all four distance 
indicators in the LMC and SMC. Additionally, we include in our analysis, when 
possible, photometric data of the distance indicators in the metal poor Carina 
dwarf galaxy for which photometry was also collected during the OGLE-II 
experiment. The analysis is largely differential, free from zero point and 
extinction uncertainties. 

The main conclusion is that the distance scales resulting from all four 
stellar distance indicators are fully consistent with each other. 
Therefore the distance 
scale problem is not a problem resulting from different distances yielded by 
distance indicators, but rather a problem of the proper zero point of this 
common distance scale. All four stellar distance indicators have to be treated 
as an {\it ensemble}, and any determination of the zero point for one of them 
must predict reasonable luminosities of the others. This puts strong 
constraints on many proposed calibrations of distance indicators. For example, 
very bright calibrations of Cepheids or RR~Lyr can practically be ruled out. 

At present, the most likely calibration of the zero point of the common 
distance scale, which would be consistent with observations of all four 
distance indicators, is that resulting from faint calibration of RR~Lyr stars 
or calibration of the red clump stars. With the OGLE-II photometry it leads to 
the distance moduli of ${(m-M)_{\rm LMC}=18.24}$~mag, ${(m-M)_{\rm SMC}=
18.75}$ mag and ${(m-M)_{\rm CAR}=19.94}$~mag for the LMC, SMC and Carina 
dwarf galaxy, respectively. The systematic uncertainty is of the order of 
0.07~mag, while the standard deviation of four determinations of only 0.02~mag. 

We also analyze the difference of distance moduli between the SMC and LMC. The 
average value is ${\Delta(m-M)_{\rm SMC-LMC}=0.50\pm0.03}$~mag from four 
independent measurements. This very good agreement allows us to draw 
conclusions on the interstellar extinction in the Magellanic Clouds. Our 
photometric data also provide constraints on the properties of red clump 
stars. 

Finally, the differential comparison of brightness of Cepheids with brightness 
of other distance indicators in the Magellanic Clouds, and preliminary 
observations of Cepheids in the IC1613 galaxy indicate no dependence of the 
zero point of Cepheid Period--Luminosity relation on metallicity.}{Magellanic 
Clouds -- Galaxies: distances and redshifts -- distance scale}
\vspace*{12pt}
\Section{Introduction}
The distance to the Large Magellanic Cloud is one of the most important 
distances in astrophysics because it sets the zero point of the extragalactic 
distance scale. Both Magellanic Clouds harbor large populations of "major" 
stellar distance indicators, namely Cepheid and RR~Lyr variable stars, red 
clump giants and red giants reaching the tip of the red giant branch (TRGB). 
Therefore they are an ideal place for testing their properties and calibrating 
their brightness. Unfortunately, the distance to the LMC has been a subject of 
controversy for a long time. After release of the Hipparcos catalog the 
situation became even more unclear. Large number of papers appeared in 
literature each claiming determination of the distance to the LMC with high 
precision and accuracy (see for example the review of Gibson 2000). The range of 
determined distance moduli, ${(m-M)_{\rm LMC}}$, increased to 18.1--18.7~mag. 
What worse, even determinations for the same distance indicator considerably 
differed \eg for Cepheids from  18.3~mag (Luri \etal 1998) to 18.7~mag (Feast 
and Catchpole 1997). However, the general feeling is that the Cepheid variables 
and TRGB stars yield longer distance moduli of ${(m-M)_{\rm LMC}\approx
18.5}$~mag or more ("long" distance scale) while RR~Lyr and the best 
calibrated red clump stars give ${(m-M)_{\rm LMC}\approx18.3}$~mag ("short" 
distance scale). Nevertheless, the opposite conclusions can also be found 
in literature. The distance modulus to the LMC seemed to diverge rather than 
converge to one well established value. 

Part of this ambiguity could be attributed to the poor photometric coverage of the 
Magellanic Clouds in the past. Fortunately, the situation in this field 
considerably changed when microlensing survey programs began regular 
photometric monitoring of these galaxies. For example the OGLE-II microlensing 
project covered photometrically large areas of both Magellanic Clouds in 
the standard {\it BVI}-bands (Udalski \etal 1998b, Udalski \etal 2000) 
very suitable for 
studying stellar population properties in these galaxies. 

In this paper we attempt to clear up the problem of the distance scale 
resulting from stellar distance indicators. We summarize photometric 
properties of four "major" stellar standard candles observed in the Magellanic 
Clouds and additionally in the Carina dwarf galaxy. The basic properties of 
Cepheids, RR~Lyr and red clump stars were already presented in the earlier 
OGLE papers. Here we complete the sample of "major" distance indicators by 
presenting determination of brightness of TRGB stars. 

Extensive and homogeneous photometric material allows us to study the 
photometric properties of all four distance indicators simultaneously. We 
present a differential analysis --  comparison of ratios of brightness of all 
combinations of standard candles in each object. It is free from 
extinction and other systematic uncertainties. Similar approach for selected 
standard candles was already presented by Udalski (1998a), reanalyzed later by 
Popowski (2000), and Bersier (2000). Our study clearly indicates that the 
distance scales resulting from all four standard candles are remarkably 
consistent. We constrain the possible zero point of this common distance scale 
and derive the most likely calibrations. 

We also perform additional test by calculating the difference of distance 
moduli between the SMC and LMC. We find that all determinations are in a very 
good agreement with the standard deviation as low as ${\pm0.03}$~mag from four 
independent measurements, confirming again consistency of distance scales of 
all four distance indicators. Based on this result we may further constrain 
some photometric properties of the stellar standard candles. 

\vspace*{12pt}
\Section{Observational Data}
Observational data of the Magellanic Clouds presented in this paper were 
collected during the second phase of the OGLE microlensing search with the 
1.3-m Warsaw telescope at the Las Campanas Observatory, Chile, which is 
operated by the Carnegie Institution of Washington. The telescope was equipped 
with the "first generation" camera with a SITe ${2048\times2048}$ CCD detector 
working in drift-scan mode. The pixel size was 24~$\mu$m giving the 0.417 
arcsec/pixel scale. Observations were performed in the "slow" reading mode of 
the CCD detector with the gain 3.8~e$^-$/ADU and readout noise of about 
5.4~e$^-$. Details of the instrumentation setup can be found in Udalski, 
Kubiak and Szyma{\'n}ski (1997). 

Observations covered significant part of the central regions of both 
Magellanic Clouds. Practically the entire bars of these galaxies were covered. 
More than 4.5 square degrees (21 ${14.2\times57}$ arcmin driftscan fields) in 
the LMC were monitored regularly from January 1997 through May 2000. In the 
SMC about 2.4 square degrees (11 fields) were observed from June 1997 through 
March 2000. Additional fields in the North-West part of the LMC were monitored 
on 13 nights between November 1998 and January 1999. Collected {\it BVI} data 
were reduced to the standard system. Accuracy of transformation to the 
standard system was about 0.01--0.02~mag. The photometric data of the SMC were 
used to construct the {\it BVI} photometric maps of the SMC (Udalski \etal 
1998b). The reader is referred to that paper for more details about methods of 
data reduction, tests of quality of photometric data, astrometry, location of 
the observed fields etc. Quality of the LMC data is similar and it is 
described in the {\it BVI} photometric maps of the LMC (Udalski \etal 2000). 

The photometric data of the Carina dwarf galaxy were collected during the 1998 and 
1999 observing seasons. Part of this observing material, methods of reductions 
etc.\ were described in Udalski (1998a). Altogether 45 frames in the {\it V} 
and 49 frames in the {\it I}-band were collected. The accuracy of the zero points 
of absolute photometry is also about 0.02~mag. 

\Section{Distance Indicators in the Magellanic Clouds and Carina Dwarf Galaxy}
In the following Subsections we describe the OGLE-II photometric data 
of four "major" stellar distance indicators in the Magellanic Clouds most 
often used for the distance determination to these galaxies. 

\Subsection{Cepheid Variable Stars}
The photometric data of about 3300 Magellanic Cloud Cepheids collected during 
the OGLE-II microlensing experiment are described in great detail in the 
series of catalogs: Udalski \etal (1999b, 1999c). Period--Luminosity and 
Period--Luminosity--Color relations were analyzed in Udalski \etal (1999a). It 
should be stressed that after the latter paper was released, the OGLE 
photometry calibration was slightly revised (see Udalski \etal 2000). 
Therefore the revised coefficients of the ${P{-}L}$ and ${P{-}L{-}C}$ 
relations for Cepheids are slightly different than those published in Udalski 
\etal (1999a). The full set of the updated coefficients, as well as {\it BVI} 
photometry of individual objects can be found in the OGLE Internet 
archive.\footnote{{\it http://www.astrouw.edu.pl/\~{}ogle} or {\it 
http://bulge.princeton.edu/\~{}ogle}} Below we only provide the $P-L$ 
relations for {\it VI}-bands and $W_I$ index for fundamental mode Cepheids we 
use in further studies: 
\begin{eqnarray}
V_0=(-2.775\pm0.031)\times\log P + 17.066\pm0.021\\
I_0=(-2.977\pm0.021)\times\log P + 16.593\pm0.014\\
W_I=(-3.300\pm0.011)\times\log P + 15.868\pm0.008
\end{eqnarray}
for the LMC, and

\vskip-10pt
\begin{eqnarray}
V_0=(-2.775\pm0.031)\times\log P + 17.635\pm0.031\\
I_0=(-2.977\pm0.021)\times\log P + 17.149\pm0.025\\
W_I=(-3.300\pm0.011)\times\log P + 16.381\pm0.016 
\end{eqnarray}
\vskip10pt
\noindent
for the SMC. We assumed here the same universal (LMC) slopes of the ${P{-}L}$ 
relations in both Magellanic Clouds (Udalski \etal 1999a). The dereddened 
magnitudes of Cepheids of the period ${P=10}$~days with statistical errors 
in all these bands in the LMC and SMC are listed in Table~1. In the case of 
$W_I$ index we included uncertainties of the zero points of {\it VI} photometry. 
We used extinction maps described in Udalski \etal (1999b) and Udalski \etal 
(1999c) to deredden the LMC and SMC data. The mean reddening in the observed 
fields is ${\langle E(B-V)\rangle=0.143}$~mag and ${\langle E(B-V)\rangle=
0.087}$~mag for the LMC and SMC fields, respectively. It should be stressed 
that the same reddening was used for all four distance indicators, so unless 
the reddening is different for different groups of stars (see Section~4.5) the 
extinction scale is the same for all analyzed distance indicators. We did not 
apply any correction for the possible population effects because the possible 
dependence of Cepheid brightness on metallicity is poorly constrained, if 
present at all (see Section~4.1). 

\MakeTable{cccc}{12.5cm}{Photometry of "major" stellar standard candles}
{\hline
\noalign{\vskip3pt}
& CARINA & SMC & LMC\\
\hline
\noalign{\vskip5pt}
CEPHEIDS ($P=10^{\rm d}$) & & & \\
\noalign{\vskip3pt}
$\langle V_0^{\rm C}\rangle$ & -- & $14.86\pm0.03$ & $14.29\pm0.02$ \\
\noalign{\vskip2pt}
$\langle I_0^{\rm C}\rangle$ & -- & $14.17\pm0.03$ & $13.62\pm0.02$ \\
\noalign{\vskip2pt}
$\langle W_I^{\rm C}\rangle$ & -- & $13.08\pm0.04$ & $12.57\pm0.04$ \\
\noalign{\vskip5pt}
\hline
\noalign{\vskip5pt}
RR~LYR & & & \\
\noalign{\vskip3pt}
$\langle V_0^{\rm RR}\rangle$ & $20.50\pm0.02$ & $19.42\pm0.01$ & $18.91\pm0.01$ \\
\noalign{\vskip2pt}
$\langle V_0^{\rm RR}\rangle_{{\rm [Fe/H]}_{\rm LMC}}$ & $20.61\pm0.03$ & $19.44\pm0.02$ & $18.91\pm0.01$ \\
\noalign{\vskip5pt}
\hline
\noalign{\vskip5pt}
TRGB & & & \\
\noalign{\vskip3pt}
$\langle I_0^{\rm TRGB}\rangle$ & $16.03\pm0.05$ & $14.83\pm0.02$ & $14.33\pm0.02$\\
\noalign{\vskip5pt}
\hline
\noalign{\vskip5pt}
RED CLUMP & & & \\
\noalign{\vskip3pt}
$\langle I_0^{\rm RC}\rangle$ & $19.46\pm0.02$ & $18.35\pm0.01$ & $17.97\pm0.01$ \\
\noalign{\vskip2pt}
$\langle I_0^{\rm RC}\rangle_{{\rm [Fe/H]}_{\rm LMC}}$ & $19.67\pm0.06$ & $18.42\pm0.03$ & $17.97\pm0.01$ \\
\noalign{\vskip5pt}
\hline}

\Subsection{RR~Lyrae Variable Stars} 
Photometric properties of about 6000 RR~Lyr variable stars detected in the LMC 
and about 430 from the SMC are described in detail in Udalski \etal (2000, in 
preparation). The mean, extinction free magnitudes of this distance 
indicator in the LMC and SMC are listed in Table~1. The applied extinction 
correction was based on the same maps as for Cepheids. 

33 RR~Lyr stars were detected in the observed field of the Carina dwarf 
galaxy. We determined the mean {\it V}-band brightness of these objects in the 
same manner as that of RR~Lyr stars from the Magellanic Clouds. The final, 
dereddened value with statistical error is listed in Table~1. Reddening of 
${E(B-V)=0.06}$ (${A_V{=}0.19}$~mag) was assumed based on Mighell (1997) 
determination of $E(V{-}I){=}0.08$~mag and Schlegel, Finkbeiner and Davis 
(1998) extinction maps. The final magnitude is in very good agreement with the 
earlier determination of RR~Lyr brightness in the Carina dwarf galaxy (Udalski 1998a) based on 
much smaller observing material. 

It is well known that the {\it V}-band brightness of RR~Lyr stars is not a perfect 
brightness reference -- it depends on metallicity of stars. Because 
metallicity of RR~Lyr stars is different in all three analyzed objects one has to 
correct for that effect. Unfortunately, the mean metallicity of RR~Lyr stars in the 
analyzed galaxies is rather poorly known, because of the lack of good 
spectroscopic determinations for significant samples of these objects. The 
mean metallicity of RR~Lyr stars in the LMC seems to be about ${\rm 
[Fe/H]=-1.6}$~dex (Alcock \etal 1996, Clementini \etal 2000). In the SMC the 
mean metallicity is somewhat lower: ${\rm [Fe/H]\approx-1.7}$~dex (see Udalski 
1998a). Metallicity of RR~Lyr stars in the Carina dwarf galaxy is also poorly known. 
It is likely to be around ${\rm [Fe/H]}\approx-2.2$~dex (Smecker-Hane \etal 
1994). 

The slope of the brightness--metallicity relation for RR~Lyr stars was also a 
subject of dispute. However, it seems now to be accepted that it is of about 
0.2~mag/dex. We assume it to be equal to 0.18~mag/dex as in Udalski (1998a): 
$$V_0^{\rm RR}=(0.18\pm0.04)\times({\rm [Fe/H]}+1.6)+{\rm const}\eqno{(7)}$$

RR~Lyr stars in metal poor environment are brighter. Therefore we added 
small corrections of 0.02 mag and 0.11 mag, resulting from differences of 
metallicity to the mean dereddened magnitudes of the SMC and Carina 
dwarf galaxy RR~Lyr stars, 
bringing them down, in this manner, to the luminosity of the LMC RR~Lyr stars 
($\rm [Fe/H]=-1.6$~dex), ${\langle V_0^{\rm RR}\rangle_{{\rm [Fe/H]}_{\rm 
LMC}}}$. Luminosity at that metallicity is most often used in calibrations of 
RR~Lyr stars. Results are also listed in Table~1. The errors of these 
quantities are larger due to uncertainty of calibration (Eq.~7). 

\Subsection{Red Clump Stars} 
Red clump is one of the most prominent features in the color-magnitude
diagrams (CMDs) of the 
Magellanic Clouds (\eg Udalski \etal 2000). The mean photometry of the red clump 
stars in a few lines of sight in the LMC halo where the reddening is small and 
well constrained by Schlegel \etal (1998) extinction maps was presented in 
Udalski (2000). The mean dereddened magnitude ${\langle I_0\rangle=17.94}$~mag 
was found in these fields. 

The mean magnitudes of the red clump stars in the Magellanic Clouds were 
determined in each of the OGLE-II fields based on {\it BVI} maps of these 
galaxies (Udalski \etal 1998b -- SMC and Udalski \etal 2000 -- LMC). In total 
about 1.3 million and 350~000 red giants were used 
in the LMC and SMC, respectively. The method of determination was 
described in Paczy{\'n}ski and Stanek (1998) or Udalski \etal (1998a) and 
Stanek, Zaritsky and Harris (1998) where the first determinations of the red 
clump distance to the Magellanic Clouds were presented, based, however, on 
somewhat overestimated interstellar extinction. 

The mean, extinction free {\it I}-band magnitudes of red clump stars in the 
LMC and SMC were obtained using the same maps of interstellar extinction 
as for Cepheids and the remaining distance indicators. 
They are listed with statistical errors in Table~1. The value for the LMC is 
very similar to the mean magnitude derived in the LMC halo fields. 

The mean derredened {\it I}-band magnitude of red clump stars in the Carina 
dwarf galaxy was already determined in Udalski (1998a). However, we 
recalculated it based on our entire observing material collected during the 
OGLE project. We also used more extensive photometric calibrations than 
before. The new value (Table~1) is about 0.02~mag fainter than in Udalski (1998a). 

The mean brightness of red clump stars is also dependent on population effects 
and therefore some corrections are necessary to be able to compare the 
brightness of red clump stars from different environments. The dependence of 
the mean brightness on metallicity was originally studied by Udalski (1998a), 
and later reanalyzed by Popowski (2000). Calibration based on Hipparcos stars 
with precise spectroscopic metallicities was presented in Udalski (2000). All 
these determinations seem to suggest modest dependence on metallicity: 
$$I_0^{\rm RC}=(0.14\pm0.04)\times({\rm [Fe/H]}+0.5)+{\rm const}\eqno{(8)}$$
what is also in relatively good agreement with theoretical modeling
(Fig.~1, Girardi and Salaris 2000).

The mean metallicity of red giants in the Magellanic Clouds is also poorly 
known. It seems that the mean metallicity of the LMC red giants is $\rm 
[Fe/H]\approx-0.5$~dex (Bica \etal 1998, Olszewski \etal 1991, Cole, 
Smecker-Hane and Gallagher 2000). The mean metallicity of the SMC is about 
0.5~dex lower what can be deduced from the mean metallicity of intermediate 
age clusters (Udalski 1998b). The mean metallicity of the Carina red giants is 
${\rm [Fe/H]\approx-2.0}$~dex based on spectroscopic determinations for 52 stars 
(Smecker-Hane \etal 1999). 

We applied corrections of 0.07~mag and 0.21~mag for the SMC and Carina, 
respectively, to bring the mean luminosity of the red clump stars to the LMC 
level, ${\langle I_0^{\rm RC}\rangle_{{\rm [Fe/H]}_{\rm LMC}}}$. Red clump 
magnitudes corrected for metallicity differences with errors including 
uncertainty of calibration (Eq.~8) are also listed in Table~1. 

Theoretical modeling of red clump seems to indicate that its mean magnitude 
also depends on the age of red clump stars. On the other hand observations of 
star clusters in the LMC and SMC do not show any evident variations larger 
than 0.05~mag with age in the age range of 2--10~Gyr (Udalski 1998b). 
Therefore, we do not find necessary to apply any age related correction of the 
red clump magnitude at this stage. We will discuss this assumption later in 
Section~4.6. 

\Subsection{Tip of the Red Giant Branch}
The mean extinction free {\it I}-band magnitudes of the tip of the red giant 
branch (TRGB) were determined based on {\it BVI} maps of the Magellanic 
Clouds. First, for each OGLE field interstellar extinction 
correction according to the OGLE extinction map in the Magellanic Clouds was 
applied. Then the stars from the range of ${12<I_0<17}$~mag and ${0.7<(V-
I)_0<3.0}$~mag were extracted. Figs.~1--3 show the upper part of the red giant 
branch for two fields in the LMC and the SMC. The bar field of the LMC 
includes all OGLE bar fields except for the eastern fields LMC$\_$SC1 and 
LMC$\_$SC16--20 where extinction is larger and non-uniform in the field. The 
NW field of the LMC includes five OGLE fields (LMC$\_$SC22--26) which were 
observed only a few times for CMDs. These fields overlap with the area of the 
LMC observed by Zaritsky, Harris and Thompson (1997) so the photometry can be directly 
compared. The extinction there is smaller: the average values are ${E(B-
V)=0.110}$~mag and $E(B-V)=0.135$~mag in the NW field and bar field of the LMC, 
respectively. 
\setcounter{figure}{3}
\begin{figure}[p]
\vglue-1cm
\includegraphics[bb=30 45 520 445,width=12cm]{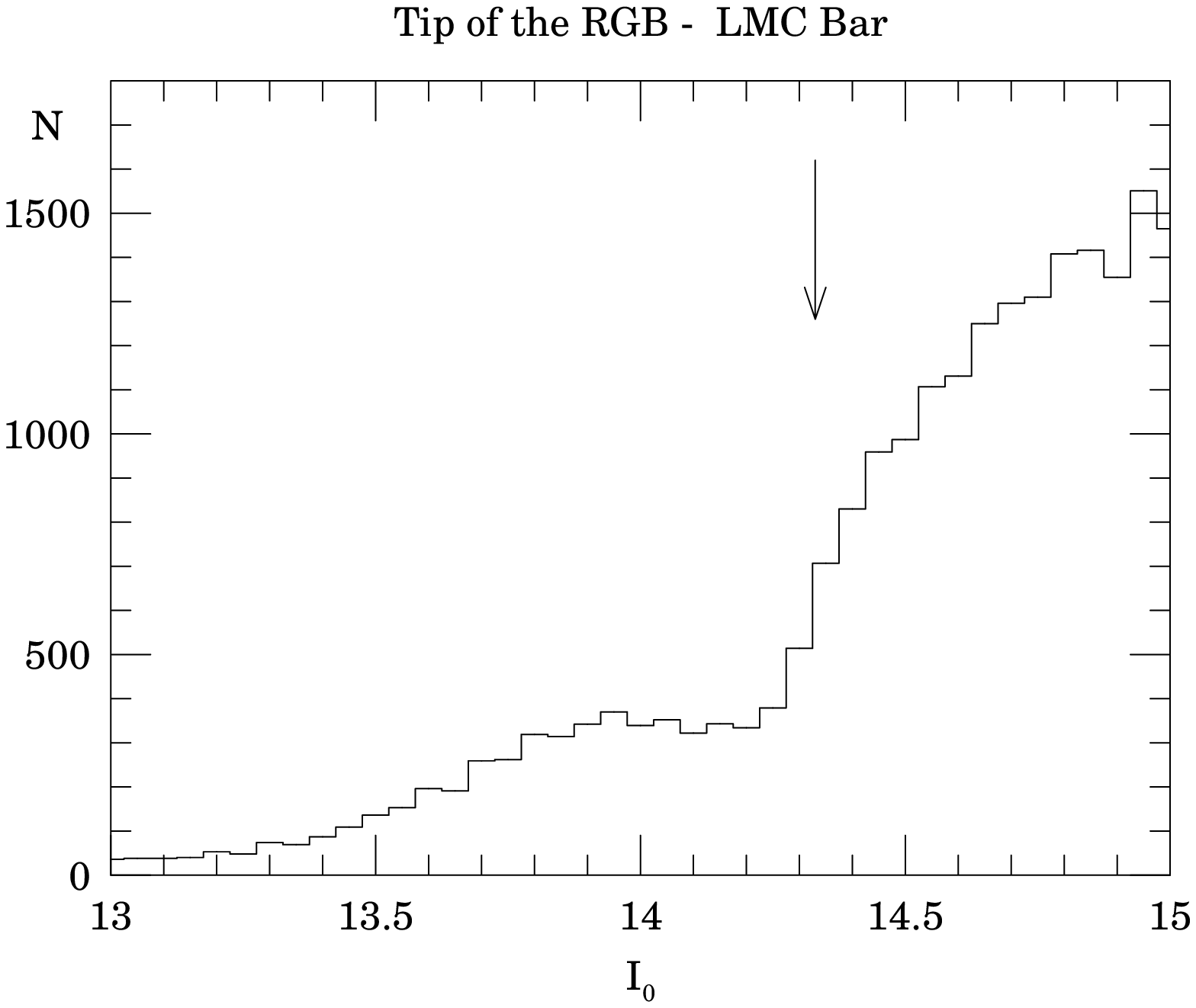}
\vskip3pt
\FigCap{Histogram of brightness of stars in the upper part of red giant
branch in the LMC bar field. Arrow marks the TRGB magnitude.}
\includegraphics[bb=30 45 520 445,width=12cm]{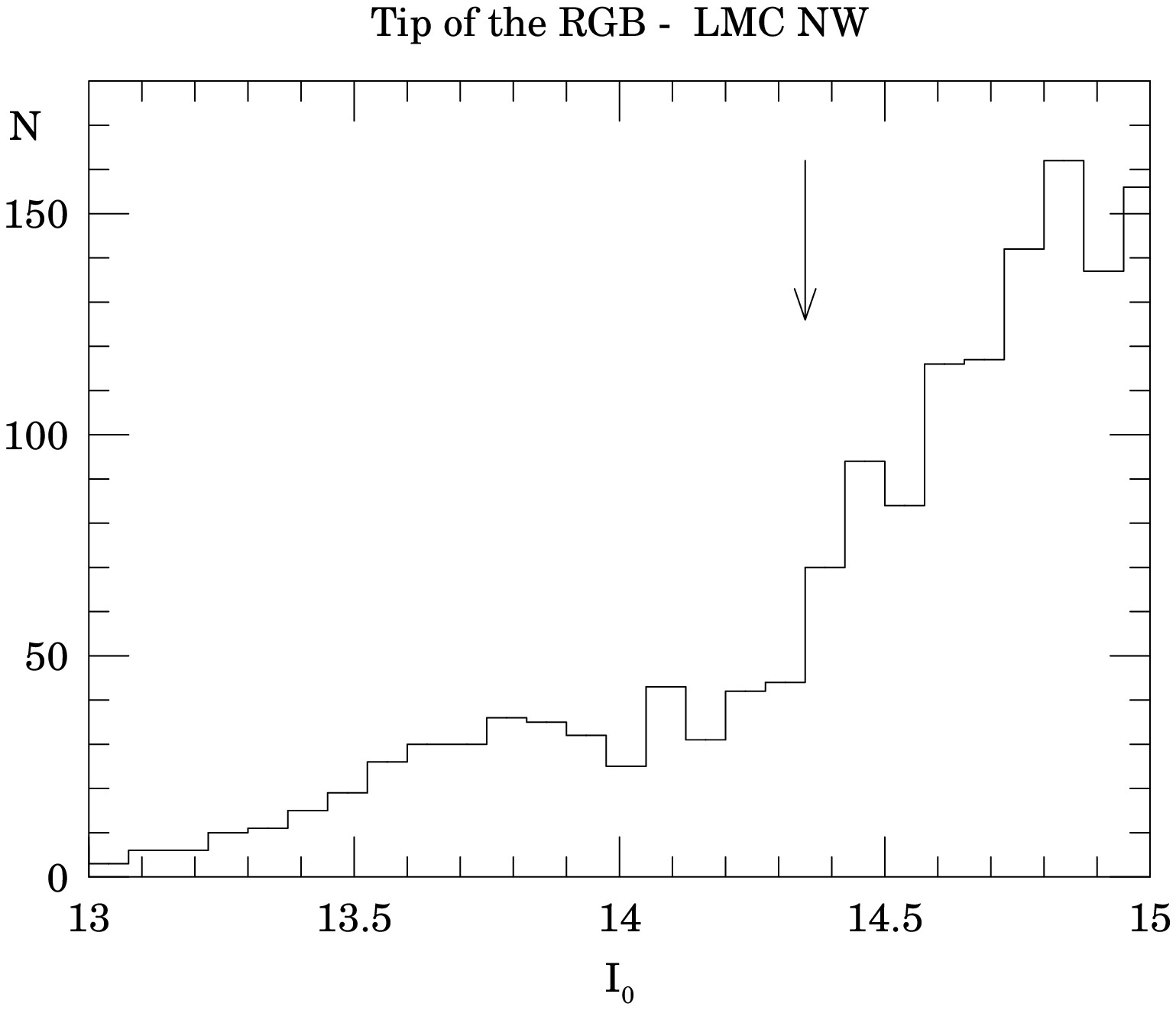}
\vskip3pt
\FigCap{Same as in Fig.~4 for the LMC NW field.} 
\end{figure}
\begin{figure}[htb]
\vglue-6mm
\includegraphics[bb=30 45 520 445,width=12cm]{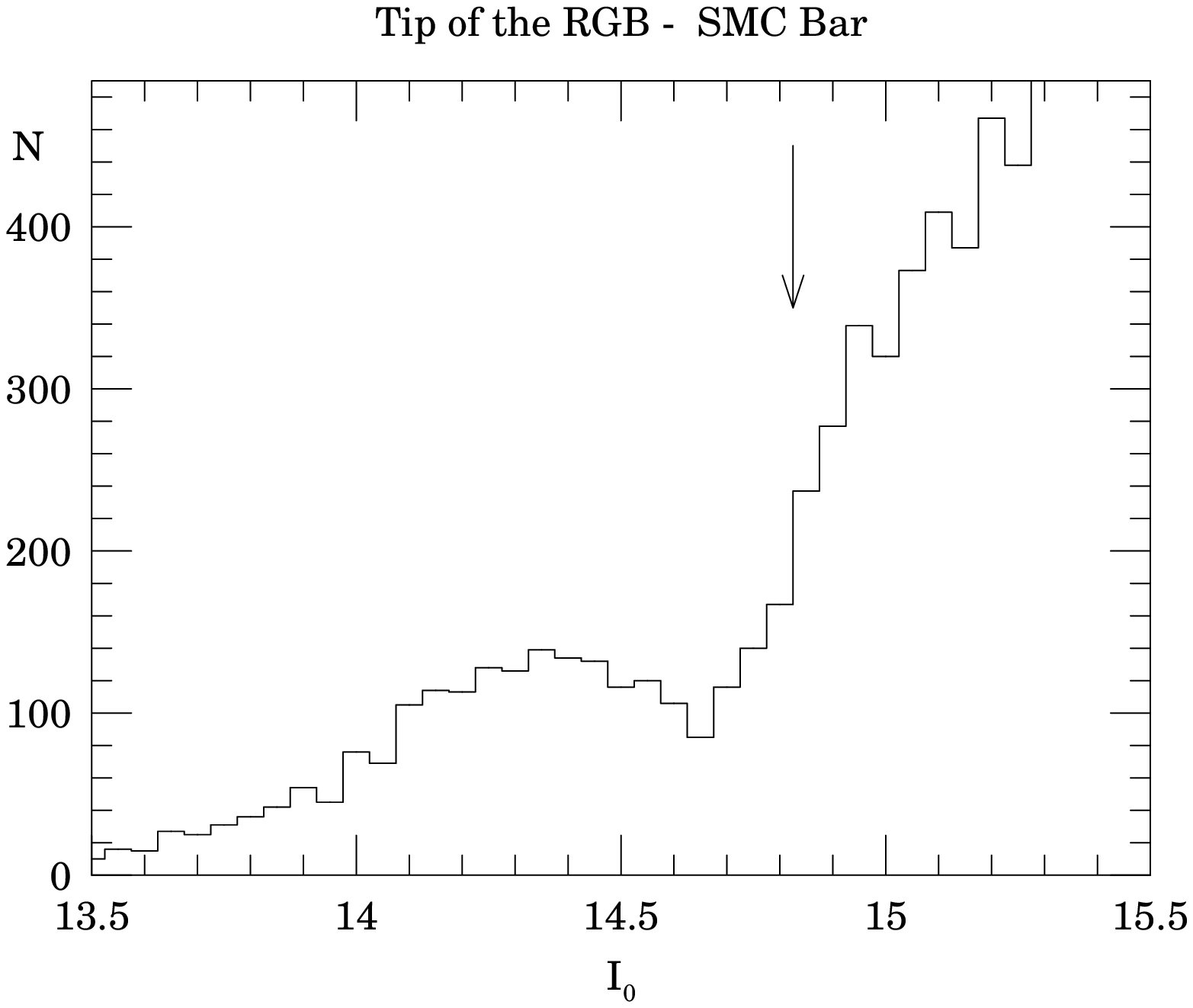}
\vskip3pt
\FigCap{Same as in Fig.~4 for the SMC field.}
\end{figure}

Figs.~1--3 show that the tip of RGB can easily be distinguished in the CMDs of 
the LMC and SMC from the remaining stars like asymptotic branch stars etc. Its 
edge is so sharp that the magnitude of TRGB can be determined with a 
ruler. Nevertheless, we constructed the histograms of luminosity function of 
stars from the RGB which are presented in Figs.~4--6. Thick solid 
line in Figs.~1--3 is the boundary of the region used for construction of 
the histograms. About 215~000, 15~000 and 57~000 stars were used in the LMC bar, 
NW and SMC fields, respectively. 

Figs.~4 and 5 indicate that the magnitude of the TRGB 
and its statistical error are equal to ${\langle I_0^{\rm TRGB}\rangle=14.33\pm0.02}$~mag 
and $14.35\pm0.05$~mag in the bar and NW LMC fields, respectively. In the 
SMC it is equal to $\langle I_0^{\rm TRGB}\rangle=14.83\pm0.02$~mag (Table~1). 
It is very encouraging that 
for both LMC fields, calibrated independently and with different reddening, 
the $\langle I_0^{\rm TRGB}\rangle$ values are almost identical. 
Determination for the bar field is much more reliable because of order
of magnitude better statistic of the stars.

The brightness of the TRGB is believed to be constant in the {\it I}-band to 
better than ${\pm0.1}$~mag for metal poor populations with metallicity lower 
than $\rm [Fe/H]\approx-0.7$~dex and older stars (${>4}$~Gyr) (\eg Lee, Freedman 
and Madore 1993a, Castellani \etal 2000a). For younger or metal richer stars the TRGB brightness 
is smaller. Metallicity of the intermediate age population in the LMC is 
dangerously close to that limit, but the LMC also contains quite numerous 
population of older stars which should fulfill the above limits and therefore 
the derived TRGB magnitude should be correct. This problem can easily be 
tested by comparison of the TRGB magnitudes in the Magellanic Clouds, \ie 
determining the difference of distance moduli between the SMC and LMC. 
According to our determinations it is equal to 0.50~mag. 
This is in excellent agreement with the difference indicated 
by other distance indicators (see Section 4.4). We are therefore confident 
that the TRGB magnitude in the LMC corresponds to the older, metal poorer 
population and is not biased by younger, metal richer stars. 

Unfortunately, the field in the Carina dwarf galaxy observed during the OGLE 
project was too small to allow precise determination of the TRGB magnitude in 
this galaxy. Therefore we used the TRGB magnitude provided by Smecker-Hane 
\etal (1994): ${\langle I^{\rm TRGB}\rangle=16.15\pm0.05}$~mag. Although a 
shift of the photometric zero points between the Smecker-Hane \etal (1994) 
and our data sets is possible, it must be smaller than about 0.03~mag as 
indicated by the location of the red clump. The dereddened TRGB magnitude was 
found by applying the same reddening correction (${E(B-V)=0.06}$~mag, 
${A_I=0.12}$~mag), as for other distance indicators in the Carina dwarf 
galaxy. 

\Section{Discussion}
Photometry of the Magellanic Clouds collected during the OGLE-II microlensing 
project provides a unique opportunity to study all four "major" stellar 
distance indicators in different environments and to analyze their properties 
as standard candles with the same homogeneous observational dataset (Table~1). So 
far the vast majority of comparisons of different distance indicators were 
based on the final distance determinations. However, such comparisons are 
usually meaningless because the observations were usually collected in 
different regions of the Magellanic Clouds, by different observers  and 
different interstellar reddening corrections were used. The number of possible 
systematic errors is large and difficult to estimate. With the huge OGLE-II 
photometric dataset we are in position to analyze for the first time all four 
distance indicators simultaneously. 
\MakeTable{cccc}{12.5cm}{Ratio of brightness of "major" stellar standard 
candles} 
{\hline
\noalign{\vskip4pt}
& CARINA & SMC & LMC\\
\hline
\noalign{\vskip4pt}
RR~LYR -- CEPHEIDS & & & \\
\noalign{\vskip3pt}
$ \langle V_0^{\rm RR}\rangle_{{\rm [Fe/H]}_{\rm LMC}}-\langle V_0^{\rm C}\rangle$
& -- & $4.58\pm0.04$ & $4.62\pm0.03$ \\
\noalign{\vskip2pt}
$ \langle V_0^{\rm RR}\rangle_{{\rm [Fe/H]}_{\rm LMC}}-\langle W_I^{\rm C}\rangle$
& -- & $6.36\pm0.04$ & $6.34\pm0.03$ \\
\noalign{\vskip4pt}
\hline
\noalign{\vskip4pt}
TRGB -- CEPHEIDS & & & \\
\noalign{\vskip3pt}
$ \langle I_0^{\rm TRGB}\rangle-\langle I_0^{\rm C}\rangle$
& -- & $0.66\pm0.04$ & $0.71\pm0.03$ \\
\noalign{\vskip2pt} 
$ \langle I_0^{\rm TRGB}\rangle-\langle W_I^{\rm C}\rangle$
& -- & $1.75\pm0.04$ & $1.76\pm0.04$ \\
\noalign{\vskip4pt}
\hline
\noalign{\vskip4pt}
RED CLUMP -- CEPHEIDS & & & \\
\noalign{\vskip3pt}
$ \langle I_0^{\rm RC}\rangle_{{\rm [Fe/H]}_{\rm LMC}}-\langle I_0^{\rm C}\rangle$
& -- & $4.25\pm0.04$ & $4.35\pm0.03$ \\
\noalign{\vskip3pt} 
$ \langle I_0^{\rm RC}\rangle_{{\rm [Fe/H]}_{\rm LMC}}-\langle W_I^{\rm C}\rangle$
& -- & $5.34\pm0.05$ & $5.40\pm0.04$ \\
\noalign{\vskip4pt}
\hline
\noalign{\vskip4pt}
RED CLUMP -- TRGB & & & \\
\noalign{\vskip4pt}
$ \langle I_0^{\rm RC}\rangle_{{\rm [Fe/H]}_{\rm LMC}}-\langle I_0^{\rm TRGB}\rangle$
& $3.64\pm0.08$ & $3.59\pm0.04$ & $3.64\pm0.03$ \\
\noalign{\vskip4pt}
\hline
\noalign{\vskip4pt}
RR~LYR -- TRGB & & & \\
\noalign{\vskip3pt}
$ \langle V_0^{\rm RR}\rangle_{{\rm [Fe/H]}_{\rm LMC}}-\langle I_0^{\rm TRGB}\rangle$
& $4.58\pm0.07$ & $4.61\pm0.04$ & $4.58\pm0.04$ \\
\noalign{\vskip4pt}
\hline
\noalign{\vskip4pt}
RR~LYR -- RED CLUMP & & & \\
\noalign{\vskip3pt}
$ \langle V_0^{\rm RR}\rangle_{{\rm [Fe/H]}_{\rm LMC}}-\langle I_0^{\rm RC}\rangle_{{\rm [Fe/H]}_{\rm
LMC}}$ & $0.94\pm0.07$ & $1.02\pm0.05$ & $0.94\pm0.03$ \\
\noalign{\vskip4pt}
\hline}

\Subsection{Ratio of Brightness of Distance Indicators}
To compare brightness of "major" distance indicators in the Magellanic Clouds 
and Carina dwarf galaxy we calculated the ratios of brightness of all possible 
pairs of distance indicators. It should be stressed that our analysis, as long 
as the brightness in the same passbands is compared, is fully differential as it 
is performed on the same dataset. Therefore all systematic errors, like 
uncertainty of the zero point of photometry, uncertainty of interstellar 
extinction (as long as different standard candles are equally reddened -- see 
Section~4.5) etc., cancel. 

In Table~2 we list the ratios of brightness (differences of magnitudes) of all 
possible pairs of four "major" stellar standard candles. For RR~Lyr and red 
clump stars we used the magnitudes corrected for metallicity effects, \ie 
lumino\-sities which the stars would have if they were of the LMC metallicity. 
For Cepheid comparisons we list the ratios for {\it VI}-bands and additionally 
for extinction insensitive index $W_I$ (Udalski \etal 1999a). In the latter 
case and in all inter-color comparisons the errors are somewhat larger because 
these comparisons are not fully differential and the zero point uncertainties 
must be taken into account. 

Figs.~7 and 8 present visualization of the data contained in Table~2. Fig.~7 
shows the differences of magnitudes between the mean brightness of Cepheids 
and other standard candles in the LMC and SMC (unfortunately the Carina dwarf 
galaxy does not host Population~I Cepheids) while Fig.~8 -- for the remaining 
combinations of standard candles. The data for the Carina dwarf galaxy are 
included in Fig.~8. The abscissa in Fig.~8 represents schematically 
metallicity of the environment. For all populations of stars in these three 
galaxies the order of increasing metallicities is Carina $\rightarrow$ SMC 
$\rightarrow$ LMC and the difference of metallicity between the Carina dwarf 
galaxy and SMC is about three times larger than between the SMC and LMC. 
\begin{figure}[htb]
\centerline{\includegraphics[bb=120 315 510 710, width=10.5cm]{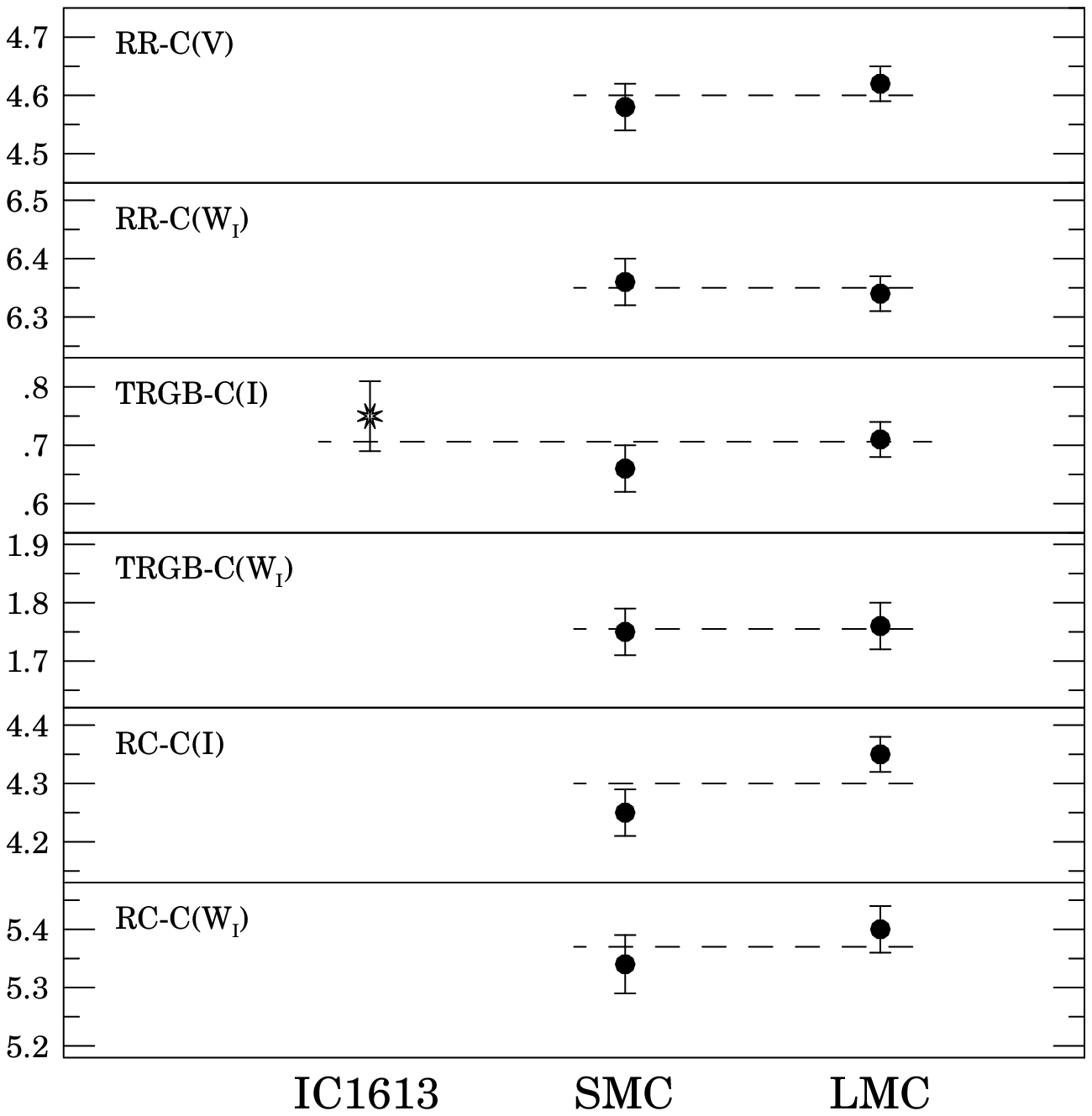}}
\vskip3pt
\FigCap{Differences of magnitudes of Cepheids and other distance indicators.}
\end{figure}
\begin{figure}[htb]
\centerline{\includegraphics[bb=125 370 505 710, width=10.5cm]{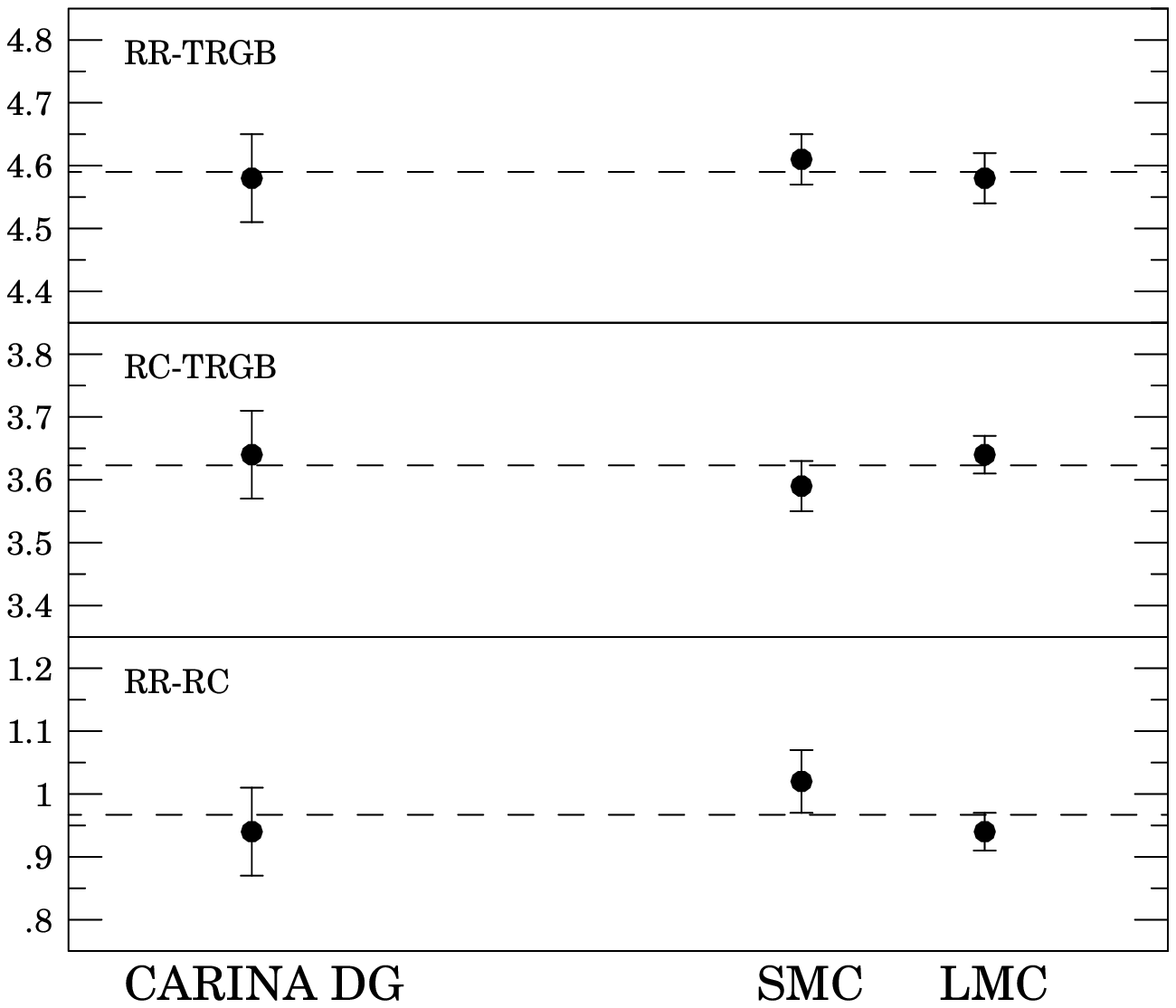}}
\vskip3pt
\FigCap{Differences of magnitudes of RR Lyr, red clump and TRGB stars.}
\end{figure}

If the standard candles were indeed good brightness references then one would 
expect that their brightness differences should be constant in all considered 
environments, \ie their magnitude differences should be parallel to the 
abscissa axes in Figs.~7 and 8. Figs.~7 and 8 clearly indicate that this is 
indeed the case (dashed line). All combinations of distance indicators present 
basically the same picture, \ie their differences of magnitude in different 
objects are constant to ${\pm0.05}$~mag in the worst case. 

Looking carefully at Fig.~7 or the Cepheid data in Table~2 one can notice a 
small systematic trend -- the differences of magnitude between the distance 
indicators and Cepheids seem to be systematically slightly larger in the LMC 
than in the SMC. This effect could result from a small dependence of Cepheid 
brightness on metallicity. Therefore to check if the metallicity effect can be 
real we observed Cepheids in the metal poor galaxy IC1613 (metallicity of 
IC1613 is smaller than that of the SMC; metallicity of Cepheids in the LMC and 
SMC is equal to ${\rm [Fe/H]=-0.3}$~dex and ${-0.7}$~dex respectively, Luck 
\etal 1998). Although this project is still in progress, preliminary analysis 
of the data collected so far indicates that the difference of the mean 
{\it I}-band brightness of Cepheids with 10~day period and TRGB stars in the 
IC1613 galaxy is equal to ${\langle M^{\rm TRGB}_I\rangle-\langle M^{\rm C}_I
\rangle=0.75\pm0.06}$~mag. Asterisk in the appropriate panel of Fig.~7 marks 
that value and the horizontal location of points corresponds to the mean 
metallicity of objects. This result confirms that the brightness of Cepheids 
does not depend on metallicity to a few hundredths of magnitude in the wide 
metallicity range of ${\approx-1<{\rm[Fe/H]}<-0.3}$~dex. More detailed 
analysis of the IC1613 data will be presented when this sub-project of the OGLE 
search is finished. 

The conclusion from our exercise is very straightforward. All four "major" 
distance indicators, namely Cepheids, RR~Lyr, red clump and TRGB stars are 
equally good standard candles. It is worth noting that these standard candles 
represent very different populations of stars from the oldest RR~Lyr stars, 
through somewhat younger TRGB population, intermediate age red clump stars and 
finally young population of classical Cepheids. The distance scales from all 
four distance indicators are fully compatible with each other, and all four 
distance indicators can be equally well used for distance determinations 
within the domain they are considered to be a standard candle. Fig.~8 strongly 
suggests that the small corrections for difference of metallicity of RR~Lyr 
and red clump stars (Eqs.~7 and 8) we applied are correct. Otherwise 
systematic deviations would be observed. It is worth noticing that the consistency 
of distances determined from the red clump and TRGB stars was already noted by 
Bersier (2000). 

\vskip15pt
\Subsection{Constraints on the Zero Point}
\vskip6pt

As we showed in Section~4.1 all four distance indicators are fully 
compatible with each other and, therefore, the distance scales determined from 
them are fully consistent. Thus, the only important problem related to the 
distance determination is a proper calibration of the zero point of this 
common distance scale. Because the brightness ratios of all four indicators 
are well constrained based on our differential analysis, all four indicators 
must be treated as an {\it ensemble}. The calibration of any of them sets the 
calibrations of all the remaining indicators. In other words any proposed 
calibration of any of the "major" stellar distance indicators must predict 
reasonable calibrations for the remaining ones. In this way we can put 
constraints on the possible calibration of the zero point of the common 
distance scale. 

First, let us briefly discuss proposed calibrations of stellar distance 
indicators. The Cepheid calibrations fall into three categories: bright, 
medium and faint. Bright calibrations predict ${M_V^{\rm C}\approx-4.2}$~mag 
for a Cepheid with the period of ${P=10}$ days (Feast and Catchpole 1997, 
Lanoix, Paturel and Garnier 1999, Groenewegen and Oudmaijer 2000) and are based on analyses of 
Hipparcos direct parallaxes of these stars. One should, however, remember that 
the parallaxes of Cepheids measured by Hipparcos are very small and have very 
large errors (only a few Cepheids has the relative error smaller than 30\%). 
Medium brightness calibrations give ${M_V^{\rm C}\approx-4.05}$~mag and are 
mostly based on Barnes--Evans method (Gieren \etal 1997) and pre-Hipparcos 
Galactic calibrations (Laney and Stobie 1994). Faint calibrations predict 
${M_V^{\rm C}\approx-3.85}$~mag based on a kind of statistical parallaxes 
method (Luri \etal 1998). Indirect calibration {\it via} Cepheids in the NGC4258 
galaxy also indicates faint brightness of Cepheids (Maoz \etal 2000). 

Calibrations of RR~Lyr stars in the post Hipparcos era were reviewed and 
summarized by Popowski and Gould (1998). This review concludes that the most 
likely absolute magnitude of RR~Lyr stars is ${M_V^{\rm RR}\approx0.70}$~mag 
at metallicity of ${\rm [Fe/H]=-1.6}$~dex. However, although the majority of 
empirical determinations seem to point to fainter calibrations of RR~Lyr, 
subdwarf fitting technique results in much higher luminosity of RR~Lyr 
stars, namely ${M_V^{\rm RR}=0.45}$~mag at ${\rm [Fe/H]=-1.6}$~dex  
(Carretta \etal 2000). The drawbacks of this method when applied to RR~Lyr calibration are 
described in Popowski and Gould (1998). Nevertheless, this calibration is 
still a possible alternative to the faint RR~Lyr calibration. 

Two kinds of calibrations are used for distance determination with TRGB stars. 
The empirical calibration based on Galactic globular clusters data (Lee \etal 
1993a) gives ${M_I^{\rm TRGB}=-4.0\pm0.1}$~mag for metallicities ${\rm [Fe/H]<-
0.7}$~dex. It is, however, based on the old RR~Lyr stars calibration 
predicting ${M_V^{\rm RR}=0.55}$~mag at ${\rm [Fe/H]=-1.6}$~dex. This is by 
about 0.1--0.15~mag brighter than the present, most likely calibration of 
RR~Lyr stars (Popowski and Gould 1998). Therefore, the correction of this order 
should be taken into account when deriving the distance with TRGB what has 
never been done. The second calibration is based on the theoretical modeling of 
TRGB stars and predicts TRGB magnitude brighter by about 0.1~mag than the 
empirical one (Salaris and Cassisi 1997). However, because the uncertainties 
related to the modeling are still rather large (Dominguez \etal 1999, Castellani 
\etal 2000b) we do not consider this calibration as a very strong alternative to 
the empirical one. 

Finally, the calibration of red clump stars is superior to calibration of any 
other standard candle what is one of the main advantages of this distance 
indicator. Because these stars are very numerous in the solar neighborhood, the 
Hipparcos sample of red clump stars with good accuracy parallaxes (relative 
error less than 10\%) and precise photometry is very large (several hundred 
objects, Paczy{\'n}ski and Stanek 1998). This made it possible to determine 
the absolute magnitude of red clump stars with high accuracy: ${M_I^{\rm RC}=-
0.23\pm0.03}$~mag for metallicity of ${\rm [Fe/H]=0.0}$~dex (Stanek and 
Garnavich 1999, Udalski 2000). Unfortunately, the local red clump stars only 
partially overlap in metallicity with the LMC red clump stars. However, the 
dependence on metallicity (Eq.~8) is rather modest so the mean {\it I}-band 
magnitude of the red clump stars in the LMC is ${M_I^{\rm RC}=-0.31\pm
0.04}$~mag. No significant dependence of the red clump magnitude on age 
was assumed here what, among others, is clearly supported by Fig.~8 
(Section~4.6). On the other hand, theoretical modeling of red clump stars 
indicates that its magnitude depends on both metallicity and age. 
Girardi and Salaris (2000) claim that the combined effect may 
reach as much as 0.20~mag in the LMC case. If true, the LMC red clump 
magnitude would be of ${M_I^{\rm RC}=-0.43}$~mag. 

Having in mind all these calibrations, less and more likely, we may constrain 
the possible zero point of the common distance scale. First, we may almost 
immediately exclude the bright calibration of Cepheids. For example, 
${M_V^{\rm C}=-4.21}$~mag and ${M_I^{\rm C}=-4.94}$~mag (Groenewegen and 
Oudmaijer 2000) for the Galactic Ce\-pheids would indicate the following magnitudes 
of the remaining distance indicators: $M_V^{\rm RR}=0.39$~mag, ${M_I^{\rm 
TRGB}=-4.23}$~mag and ${M_I^{\rm RC}=-0.64}$ mag. The mean magnitude of RR~Lyr 
stars would have to be brighter than the brightest calibration from subdwarf 
fitting. The TRGB magnitude would be much brighter than indicated by Lee \etal 
(1993a) calibration but this could be marginally consistent when correction to the TRGB 
calibration resulting from the appropriate RR~Lyr calibration (${\approx0.15}$~mag 
brighter) was applied. However, the predicted absolute magnitude of the red clump 
stars practically rules out the bright Cepheid calibration. The mean magnitude 
of the red clump stars of the LMC metallicity would have to be by about 
0.35~mag brighter than the empirical calibration or 0.2~mag brighter than the 
theoretical predictions. This large discrepancy could only vanish when the 
brightness of Cepheids would very strongly depend on metallicity (about 
1.0~mag/dex in the {\it I}-band) making the Galactic Cepheids much brighter. 
As we have shown in Section~4.1 our data do not indicate any 
significant dependence of Cepheid brightness on metallicity. Therefore this 
calibration can be ruled out with high confidence. 

Similar arguments can exclude the bright calibration of RR~Lyr stars. 
$M_V^{\rm RR}=0.45$~mag would indicate ${M_V^{\rm C}=-4.15}$~mag, 
${M_I^{\rm TRGB}=-4.14}$~mag and $M_I^{\rm RC}=-0.52$~mag. While Cepheids 
and TRGB calibrations could be consistent with the bright RR~Lyr calibration, 
the inferred magnitude of red clump stars is again too bright by at least 
0.1~mag from theoretical predictions or more than 0.2~mag from the empirical 
calibration. Because we have already corrected RR~Lyr stars magnitude for 
metallicity it would be extremely difficult to find additional 0.1--0.2~mag 
that could explain this inconsistency. Therefore the bright calibration of 
RR~Lyr also seems to be unlikely. 

On the other hand two of the already  discussed calibrations of
the distance indicators presented here, namely Hipparcos
calibration of red clump stars and faint calibration of RR~Lyr, are
consistent with all four standard candles. The
latter calibration (Popowski and Gould 1998), ${M_V^{\rm RR}=0.70}$~mag, would 
indicate: $M_V^{\rm C}=-3.90$~mag, ${M_I^{\rm TRGB}=-3.89}$~mag and 
${M_I^{\rm RC}=-0.27}$~mag. All these predictions are in good agreement with
the faint calibration of Cepheids 
(statistical parallaxes and Cepheids in NGC4258), Lee \etal (1993a) calibration 
of TRGB (corrected down by 0.15 mag due to fainter RR~Lyr brightness scale) 
and the empirical (Hipparcos) calibration of red clump stars (Udalski 2000), 
respectively. The empirical calibration of red 
clump stars of the LMC metallicity, ${M_I^{\rm RC}=-0.31}$~mag (Udalski 2000), 
leads to very similar predictions: ${M_V^{\rm RR}=0.66}$~mag, ${M_I^{\rm C}=-
4.61}$~mag and $M_I^{\rm TRGB}=-3.93$~mag. 

Very consistent results obtained with calibrations of RR~Lyr and red clump 
stars strongly suggest that in the entire range of possible calibrations those 
two are the most likely. If we adopt the mean results based on the RR~Lyr and 
red clump empirical calibrations and our differential analysis of "major" 
stellar standard candles then the following formulae for individual standard 
candles can be written:\\ 
Cepheids:
\setcounter{equation}{8}
\begin{eqnarray}
M_V=(-2.775\pm0.031)\times(\log P-1) - 3.92\pm0.05\\
M_I=(-2.977\pm0.021)\times(\log P-1) - 4.61\pm0.05\\
M_{W_I}=(-3.300\pm0.011)\times(\log P-1) - 5.67\pm0.05
\end{eqnarray}
TRGB:
$$M_I^{\rm TRGB}=-3.91\pm0.05~{\rm mag}\qquad({\rm [Fe/H]}<-0.7~{\rm dex})\eqno{(12)}$$
RR~Lyr:
$$M_V^{\rm RR}=(0.18\pm0.04)\times({\rm [Fe/H]}+1.6)+0.68\pm0.05 \eqno{(13)}$$
Red clump stars:
$$M_I^{\rm RC}=(0.14\pm0.04)\times({\rm [Fe/H]}+0.5)-0.29\pm0.05 \eqno{(14)}$$
where all zero points refer to the LMC metallicity and their errors
are estimated from uncertainty of the Hipparcos calibration of red clump
stars ($\pm0.04$~mag) and the consistency of both calibrations. It should be, however, 
remembered that our differential test cannot exclude somewhat brighter 
calibrations, \ie up to $M_V^{\rm RR}\approx0.55$~mag for RR Lyr and ${M_I^{\rm 
C}\approx-4.7}$~mag for Cepheids. They could be consistent at least with 
theoretical predictions for red clump brightness and thus cannot be fully 
ruled out. These limits go by about 0.1~mag fainter (\ie to Eqs.~9--14) if the 
empirical calibration of the red clump is considered. 

\Subsection{Distance to the LMC, SMC and Carina Dwarf Galaxy}
Eqs.~(9)--(14) can be directly applied to the dereddened magnitudes of standard 
candles provided in Table~1 to determine the distances to the analyzed 
galaxies. In this case, however, the analysis is no longer differential and 
any systematic errors can affect the results. While the zero points of the 
OGLE photometry were tested many times (\eg Udalski \etal 1998ab, Udalski 
\etal 2000) and it is unlikely that the systematic uncertainty from this 
factor is larger than 0.02~mag, the main source of uncertainty remains the 
interstellar extinction applied to the photometry to deredden the data. 
Accuracy of the zero points of extinction maps used by OGLE was discussed in 
Udalski \etal (1999bc). Also as we will show in Section~4.5 there is no 
indication that the interstellar reddening is different for younger and older 
populations as suggested by Zaritsky (1999). The mean reddening of 
${E(B-V)=0.143}$~mag and 0.087~mag in the bar of the LMC and SMC, 
respectively, seems to be consistent with all determinations in these regions 
with hot young or older stars (including the recent determination from RR~Lyr 
instability strip colors by Clementini \etal 2000, when correct boundaries of 
the strip, \ie bluer by about 0.05~mag, \eg from Walker 1998, are used). 
Therefore, it is rather unlikely that the uncertainty of $E(B-V)$ reddening is 
larger than 0.02~mag. 

Distances to the LMC, SMC and Carina dwarf galaxy calculated with Eqs.~(9) --(14) 
are listed in Table~3. As it can be expected all distances are very similar. This 
is not surprising as we already showed that all four distance scales are fully 
consistent. The mean values of distance moduli from four "major" stellar 
standard candles are equal to ${(m-M)_{\rm LMC}=18.24}$~mag, ${(m-M)_{\rm 
SMC}=18.75}$~mag and ${(m-M)_{\rm CAR}=19.94}$~mag. The systematic error is of 
the order of 0.07~mag and results mostly from the reddening and 
calibration (Eqs.~9--14) uncertainties while 
the standard deviation of these measurements is only of 0.02~mag. 
\renewcommand{\arraystretch}{0.9}
\MakeTable{cccc}{12.5cm}{Distance moduli to the Carina dwarf galaxy, SMC
and LMC}
{\hline
\noalign{\vskip3pt}
STANDARD CANDLE & CARINA & SMC & LMC\\
\hline
\noalign{\vskip5pt}
CEPHEIDS (mean {\it V,I},$W_I$) & -- & $18.77\pm0.07$ & $18.23\pm0.07$ \\
\noalign{\vskip3pt}
RR~LYR ({\it V})& $19.93\pm0.08$ & $18.76\pm0.07$ & $18.23\pm0.07$ \\
\noalign{\vskip3pt}
TRGB ({\it I})& $19.94\pm0.08$ & $18.74\pm0.08$ & $18.24\pm0.07$ \\
\noalign{\vskip3pt}
RED CLUMP ({\it I}) & $19.96\pm0.08$ & $18.71\pm0.07$ & $18.26\pm0.07$ \\ 
\noalign{\vskip3pt}
\hline}

The resulting distance to the LMC is in the range of "short" distances, \ie it 
is by about 0.25~mag smaller than the classical distance modulus to the LMC of 
${(m-M)_{\rm LMC}=18.5}$~mag. We can summarize that the OGLE photometric data 
of four "major" stellar distance indicators clearly indicate that the most 
likely zero point of the common distance scale which is consistent with all 
four standard candles is that which gives the "short" distance scale. Although 
we are not able to completely exclude calibrations up to 0.15~mag brighter, 
all the remaining ones seem to be inconsistent with at least one of the four 
distance indicators. It is worth noting at this point that the recent 
determinations of the distance to the LMC with independent, largely geometric 
standard candle, -- eclipsing binaries -- also provide strong support for the 
"short" distance scale to the LMC (${(m-M)_{\rm LMC}\approx18.3}$~mag, Guinan 
\etal 1998, Fitzpatrick \etal 2000). This is completely independent 
confirmation that the calibrations presented in this paper are the most 
likely. Also, as we already mentioned, the "short" distance scale to the LMC 
results from geometric determination of the distance to NGC4258 galaxy 
(Herrnstein \etal 1999) and photometry of Cepheids in this object (Maoz \etal 
1999). Nevertheless, the final solution of the zero point problem will 
probably have to wait until the next space astrometric missions (FAME, SIM, 
GAIA) are launched and finished and direct good quality parallaxes of large 
samples of Cepheids, RR~Lyr and TRGB stars are determined. 

\vspace*{-9pt}
\Subsection{Difference of Distance Moduli SMC--LMC} 
\vspace*{-5pt}
The differences of distance moduli between the SMC and LMC 
resulting from Cepheids, RR~Lyr, TRGB and red clump stars are listed in 
Table~4. They were derived using the extinction free luminosities listed in 
Table~1. 
\renewcommand{\arraystretch}{1}
\MakeTable{cc}{12.5cm}{Difference of distance moduli SMC--LMC}
{\hline
\noalign{\vskip3pt}
STANDARD CANDLE & $(m-M)_{\rm SMC}-(m-M)_{\rm LMC}$\\
\hline
\noalign{\vskip5pt}
CEPHEIDS ($W_I$) & $0.51\pm0.05$ \\
\noalign{\vskip3pt}
RR~LYR ({\it V})& $0.53\pm0.07$ \\
\noalign{\vskip3pt}
TRGB ({\it I}) & $0.50\pm0.05$ \\
\noalign{\vskip3pt}
RED CLUMP ({\it I}) & $0.45\pm0.05$\\
\noalign{\vskip3pt}
\hline}

It is striking that all independent determinations of ${\Delta (m-M)_{\rm SMC-
LMC}}$ are very similar. The most reliable one comes from the $W_I$ 
index for Cepheids, because those magnitudes are in principle extinction free. 
Therefore by comparison with $W_I$ Cepheid distance we may assess reliability 
of other distance indicators as well as the reddening scale used for distance 
determinations. 

The numbers from Table~4 can be affected by systematic error resulting from 
the uncertainty of extinction in both galaxies. It should, however, affect very 
similarly all determinations, except for $W_I$ index which is extinction 
independent. Very good agreement of the values from different distance 
indicators as compared to $W_I$ value indicates, however, that this error is 
probably very small (${<0.02}$~mag) and the extinction scales in both 
Magellanic Clouds are correct. The mean difference of distance 
moduli between the SMC and LMC from all four independent determinations 
(Cepheids, RR~Lyr, TRGB and red clump) is equal to 0.50~mag with very small 
standard deviation of only 0.03~mag. 

\Subsection{Interstellar Extinction in the LMC}
Very good agreement of $\Delta (m-M)_{\rm SMC-LMC}$ from all distance 
indicators provides strong constraints on many observables. First, we may 
practically rule out the possibility that the interstellar reddening is 
significantly different for different populations of stars in the LMC as 
suggested by Zaritsky (1999). Based on a new method he determined interstellar 
extinction in the {\it V}-band equal to ${A_V\approx0.4}$ mag for hot young 
stars and only ${A_V\approx0.1}$~mag for old cool population of stars in a 
field in the LMC overlapping with our NW field (Section~3.4). The simplest 
test of reality of Zaritsky (1999) hypothesis is to compare the most sound 
difference of distance moduli between the SMC and LMC, namely resulting from 
Cepheid extinction free index $W_I$ and equal to 0.51~mag, with that obtained from old 
populations under the assumption of very small extinction for these groups of 
stars. Even if we assume the extinction toward the SMC equal to zero (what is 
an obvious underestimate), $A_V=0.1$~mag in the LMC and the  mean 
magnitudes of RR~Lyr, TRGB and red clump stars  indicate 
that ${\Delta(m-M)_{\rm SMC-LMC}}$ from the old population would already be by 
about 0.1~mag smaller than the extinction free value resulting from Cepheid 
$W_I$ index. When non-zero reddening in the SMC is applied the gap 
widens. ${\Delta(m-M)_{\rm SMC-LMC}}$ from the old population converges to 
that from young Cepheids only when the interstellar extinction is the same for 
both populations and its value is close to that of the OGLE extinction maps. 

Another test of the Zaritsky (1999) hypothesis is provided by Fig.~7. Good 
agreement of differences of magnitudes of Cepheids and old population standard 
candles in the LMC and SMC under the assumption that the interstellar 
extinction is the same for all stars in both Magellanic Clouds practically 
rules out that hypothesis. Even if the extinction for young population were 
larger than for the old one, as Zaritsky suggests, it would have to be larger 
by exactly the same amount in the LMC and SMC. That would imply exactly the 
same properties and distribution of dust in both these galaxies what is 
extremely unlikely. Moreover, the interstellar extinction in the SMC is much 
smaller than in the LMC and simply there is no room for so large extinction 
difference as Zaritsky (1999) claims in the LMC. Only very unlikely conspiracy 
could mask the differences of magnitudes of Cepheids and old standard candles 
so they were exactly the same in both Magellanic Clouds. Both tests rather 
unambiguously indicate that the interstellar reddening, ${E(B-V)}$ in the 
Magellanic Clouds is to 0.01--0.02~mag similar for young and old populations 
and results of Zaritsky (1999) are probably an effect of his method of 
extinction determination for cool stars (because the derived interstellar 
extinction for hot stars, \ie those with simple and well known spectra, seems to 
be reasonable: ${E(B-V)\approx0.12}$~mag \vs 0.11~mag from the OGLE maps). 

\vspace*{-9pt}
\Subsection{Properties of the Red Clump Stars}
\vspace*{-5pt}
The mean extinction free {\it I}-band magnitude of red clump stars in the LMC 
is sometimes a subject of controversy. For example Romaniello \etal (1999) 
claim that it is by about 0.15~mag dimmer than the value presented in this paper. 
They determined ${\langle I_0^{RC}\rangle=18.12}$~mag in a field of very high 
and non-uniform extinction around SN1987A, based on HST photometry. Such a 
value can, however, be easily ruled out by comparison of differences of 
distance moduli between the SMC and LMC. Only if the extinction in the SMC 
were equal to zero, the difference of distance moduli inferred from red clump 
stars with Romaniello \etal (1999) value of $\langle I_0^{RC}\rangle$ could be 
consistent with the Cepheid extinction free $W_I$ index value 
(Table~4). With any non-zero extinction in the SMC the difference of 
distance moduli of red clump stars would be smaller -- up to 0.2~mag if the SMC extinction 
presented in this paper were used. This inconsistency makes the Romaniello 
\etal (1999) value highly unlikely. 

Observational data presented in this paper can also be used for constraining the 
possible dependence of the mean brightness of the red clump stars on age. 
Fig.~8, where comparison of the red clump brightness with other distance 
indicators is presented, clearly shows that the metallicity correction given 
by Eq.~(8) very well corrects the excess of red clump brightness. There are no 
systematic trends which could suggest other slope in Eq.~(8) and the residuals 
from a straight line, parallel to the abscissa axis, are of the order of 
${\pm0.04}$~mag only. If the excess of brightness of red clump stars in lower 
metallicity objects is indeed caused by metallicity differences then the very 
small residuals indicate that the red clump brightness is practically 
independent on age. It should be noted that on average the red clump in the 
Carina dwarf galaxy is much older than in the Magellanic Clouds. 

On the other hand theoretical modeling of red clump stars by Girardi and 
Salaris (2000) indicates that its magnitude is dependent on both metallicity 
and age. Girardi and Salaris (2000) constructed synthetic CMDs using their 
evolutionary models and assumed star formation rate (SFR) and age-metallicity 
relations (AMR) for the Magellanic Clouds and Carina dwarf galaxy to analyze 
the mean properties of red clump stars in these objects. They claim that the 
most likely assumptions produce, due to combined metallicity and age 
dependence, the red clump brighter by 0.20, 0.29 and 0.29~mag in the LMC, SMC 
and Carina dwarf galaxy, respectively, than the local red clump stars measured 
by Hipparcos. It is, however, extremely difficult to assess reality of these 
results. Qualitative comparison of the best synthetic CMDs with the empirical 
ones indicates that the simulations are rather far from reality -- in 
particular the synthetic SMC red clump (Fig.~17, Girardi and Salaris 2000) 
does not resemble at all very round and featureless red clump observed in this 
galaxy (Fig.~10; Paczy{\'n}ski \etal 1999). Also the LMC bar and Carina dwarf 
galaxy synthetic red clumps hardly resemble the empirical ones (Udalski \etal 
2000, Udalski 1998a). 

The Carina dwarf galaxy simulations are the most clear indication that the 
modeling is very far from being accurate. Looking at Fig.~1 of Girardi and 
Salaris (2000) one can find that it is practically impossible to obtain the 
red clump in Carina dwarf galaxy brighter than 0.1~mag than in the LMC when 
its age is larger than 3~Gyr, independently of SFR and AMR assumptions. Indeed, 
Girardi and Salaris (2000) obtain the red clump of only 0.09~mag 
brighter than the LMC red clump or even fainter if other SFR is assumed. 
Observations, however, indicate that it is by about 0.2~mag brighter (Fig.~8, 
Table~1). This is a large discrepancy and it probably gives a realistic 
information on accuracy of the models and synthetic CMD simulations. Modeling 
results while reasonably well reproducing qualitative features of stellar 
populations are not accurate to better than 0.1--0.2~mag (see also Castellani 
\etal 2000b and discussion in Paczy{\'n}ski \etal 1999). 

Another empirical confirmation of very weak and shallow dependence of the mean 
{\it I}-band magnitude of the red clump on age within the age range of 
2--10~Gyr provides comparison of red clump and TRGB magnitudes. Bersier (2000) 
presented such a comparison for the Magellanic Clouds, Carina and Fornax dwarf 
galaxies and found that the magnitude correction for metallicity (Eq.~8) makes 
these two distance indicators very consistent. His results are fully confirmed 
in this paper (Fig.~8). As we mentioned, small residuals indicate that 
the age dependence must be weak. Because of large brightness, the TRGB 
magnitude is relatively easy to determine in many even quite far located 
objects (similar brightness Cepheids require at least 20--30 epochs to determine 
their mean magnitudes). Therefore, we attempted to extend the comparison of 
the mean red clump and TRGB magnitudes to the Local Group galaxies with 
populations suitable for such comparisons (low metallicity, older stars for 
TRGB and intermediate age stars for red clump). Beside the Magellanic Clouds 
and Carina dwarf galaxy data presented in this paper we found in literature data for 
nine additional objects whose CMDs include both TRGB and red clump stars. We 
limited ourselves to objects in which the red clump is reasonably above the limit of photometry so 
that its mean magnitude is not affected. The data for all objects are listed 
in Table~5 in order of decreasing metallicity of the object, as well as 
${\langle M^{\rm RC}_I\rangle-\langle M^{\rm TRGB}_I\rangle}$ calculated with the 
red clump magnitude converted to the LMC metallicity (\ie with correction, 
LMC$^{\rm RC}_{\rm COR}$, listed in the fifth column in Table~5 
and obtained from Eq.~8). Unfortunately, in all 
but one cases (NGC147) we did not have the original data, so that we had to 
rely on authors' determinations or in a few cases to determine the mean 
magnitudes from CMD plots. Therefore, the presented exercise is not fully 
homogeneous. Nevertheless, the magnitudes presented in Table~5 should be 
accurate to about 0.05~mag. We usually used the same photometry to determine 
the difference of red clump and TRGB magnitudes to avoid systematic errors. In 
a few cases, however, we had to use additional source for TRGB magnitude -- 
the HST fields where photometry reached red clump stars were too small for reliable TRGB 
magnitude determination. In these cases we first verified whether both 
photometric data sets were on the same scale. 

Differences ${\langle M^{\rm RC}_I\rangle-\langle M^{\rm TRGB}_I\rangle}$, listed 
in the sixth column of Table~5, indicate that except for two of twelve objects 
they all fall in the narrow range of $3.62\pm0.08$~mag. This sample includes 
objects with younger intermediate age population like LMC and older objects -- 
Fornax, M32 (${\approx8.5}$~Gyr) or Leo~II ($\approx 9$~Gyr). One should note 
that the Girardi and Salaris (2000) modeling (their Fig.~1) predicts red clump 
fainter by about by 0.25~mag in the latter cases. Clearly, the dependence on 
age is very weak. 

Two deviating cases in Table~5 can easily be explained. In the case of Leo~A 
the vast majority of red clump population are the young stars of 0.9--1.5~Gyr 
age (Tolstoy \etal 1998). It is not surprising that they are brighter making 
the difference of red clump and TRGB magnitudes smaller. NGC147 is on the  
opposite end. The vast majority of its population is older than 10~Gyr, so 
that it is not surprising that the red clump magnitude drops, making the 
difference much larger. The NGC147 case is very similar to the NGC121 cluster 
case in the SMC (Udalski 1998b), also possessing old (${\approx12}$~Gyr) 
population and the red clump by about 0.4~mag fainter than that of younger 
clusters. Results of comparison of red clump and TRGB magnitudes in the nearby 
Local Group galaxies are in fact very consistent with results of the analysis of 
clusters in the LMC and SMC (Udalski 1998b) also indicating that in the 
2--10~Gyr range the dependence of the red clump brightness on age is very weak. 

Concluding, presently we do not find any convincing arguments on reality of 
significant dependence of the red clump brightness on age, and on the contrary 
the empirical data presented in this paper rather ~unambiguously~ seem to 
indicate that such a dependence is marginal for ages within 2--10~Gyr. This 
confirms our earlier results from analysis of the Magellanic Cloud star 
clusters (Udalski 1998b). That analysis, in particular of the SMC clusters, 
was criticized by Girardi and Salaris (2000) because of applying geometric 
corrections resulting from the spatial extent of the SMC (these corrections 
result, however, from our current picture of the SMC 
geometry). Nevertheless, we would like to point attention to the fact that at 
least two clusters of very different age from the Udalski's (1998b) sample are 
located very close to each other on the sky (Kron~3 -- 7.5~Gyr and L11 -- 
3.5~Gyr) so their photometry can be compared differentially. The mean 
magnitude of red clump in these objects is similar to $\pm0.02$~mag after 
correcting for a small metallicity difference. Girardi and Salaris (2000, 
Fig.~1) models predict the red clump brighter by about 0.15~mag in younger 
L11. The brightness is also very similar to the mean magnitude of red clump of 
field stars indicating that the clusters are located at the same distance as 
the bulk of SMC stars in that direction. This is the only pair of clusters of 
different age allowing direct comparison of red clump magnitude. Other 
comparisons, for example those of Galactic clusters (Girardi and Salaris 
2000), suffer from uncertainties of distance much larger than the geometric 
correction uncertainties in the SMC and very small number statistics of red 
clump stars. 

\begin{landscape}
\MakeTableSep{
l@{\hspace{15pt}}c@{\hspace{15pt}}c@{\hspace{9pt}}c@{\hspace{9pt}}
r@{\hspace{9pt}}l@{\hspace{9pt}}c@{\hspace{9pt}}c@{\hspace{9pt}}
}{12.5cm}{TRGB and red clump stars in nearby galaxies}
{\hline
\noalign{\vskip3pt}
\multicolumn{1}{c}{GALAXY}& TRGB & RED CLUMP & [Fe/H] & 
\multicolumn{1}{c}{LMC$^{\rm RC}_{\rm COR}$}& \multicolumn{1}{c}{RC--TRGB} & Ref. & Remarks\\
 & [mag]& [mag] & [dex] & \multicolumn{1}{c}{[mag]}&\multicolumn{1}{c}{[mag]} & & \\
\hline
\noalign{\vskip5pt}
M32 & $\langle I\rangle=20.75$ & $\langle I\rangle=24.35$ & $-0.3$ &$-0.03$~~~~
& ~~~~3.55 & (1) & $\approx 8.5$~Gyr \\
LMC & $\langle I_0\rangle=14.32$ & $\langle I_0\rangle=17.95$ & $-0.5$ &$0.00$~~~~
& ~~~~3.64 & this paper& \\
M31 (G302) Field & $\langle I\rangle=20.65$ & $\langle I\rangle=24.25$ & $-0.6$ &$0.01$~~~~
& ~~~~3.6 & (2) & \\
SMC & $\langle I_0\rangle=14.80$ & $\langle I_0\rangle=18.34$ & $-1.0$ &$0.07$~~~~
& ~~~~3.59 & this paper& \\
Fornax & $\langle I\rangle=16.65$ & $\langle I\rangle=20.24$ & $-1.0$ &$0.07$~~~~
& ~~~~3.66 & (3) & \\
Pegasus& $\langle I\rangle=20.95$ & $\langle I\rangle=24.50$ & $-1.1$ &$0.08$~~~~
& ~~~~3.65 & (4), (5) & \\
Phoenix& $\langle I\rangle=19.00$ & $\langle I\rangle=22.50$ & $-1.4$ &$0.13$~~~~
& ~~~~3.65 & (6) & \\
Leo~II & $\langle I\rangle=17.70$ & $\langle I\rangle=21.25$ & $-1.6$ &$0.15$~~~~
& ~~~~3.7 & (7), (8) & $\approx 9$~Gyr\\
Carina & $\langle I_0\rangle=16.03$ & $\langle I_0\rangle=19.46$ & $-2.0$ &$0.21$~~~~
& ~~~~3.64 & this paper, (9)\\
Leo~I & $\langle I\rangle=18.25$ & $\langle I\rangle=21.60$ & $-2.0$ &$0.21$~~~~
& ~~~~3.55 & (10) & \\
\noalign{\vskip3pt}
\hline
\noalign{\vskip3pt}
Leo~A & $\langle I\rangle=20.50$ & $\langle I\rangle=23.77$ & $-1.7$ &$0.17$~~~~
& ~~~~3.44 & (11) & $0.9-1.5$~Gyr\\
NGC147 Inner Field& $\langle I\rangle=20.31$ & $\langle I\rangle=24.02$ & $-0.9$ &$0.06$~~~~
& ~~~~3.77 & (12) & $>10$~Gyr \\
NGC147 Outer Field& $\langle I\rangle=20.31$ & $\langle I\rangle=24.06$ & $-1.0$ &$0.07$~~~~
& ~~~~3.82 & (12) & $>10$~Gyr \\
\noalign{\vskip3pt}
\hline
\noalign{\vskip3pt}
References:&&&&&&&\\
\multicolumn{8}{l}{(1) Grillmair \etal (1996), (2) Holland, Fahlman and Richer (1996), (3) Bersier (2000),}\\
\multicolumn{8}{l}{(4) Aparicio (1994), (5) Gallagher \etal (1998), (6) Mart{\'{\i}}nez-Delgado \etal (1998),}\\
\multicolumn{8}{l}{(7) Lee (1995), (8) Mighell and Rich (1996), (9) Smecker-Hane \etal (1994),}\\
\multicolumn{8}{l}{(10) Lee \etal (1993b), (11) Tolstoy \etal (1998), (12) Han \etal (1997).}\\
}
\end{landscape}

\Acknow{We would like to thank Drs.\ B.\ Paczy\'nski, K.Z.\ Stanek,  
and M.~Szyma{\'n}ski for 
comments and remarks on the paper. The paper was partly supported by the 
Polish KBN grant 2P03D00814 to A.\ Udalski and 2P03D00916 to M.\ 
Szyma{\'n}ski. Partial support for the OGLE project was provided with the NSF 
grant AST-9820314 to B.~Paczy\'nski.}

\setcounter{figure}{0}
\newpage
\begin{figure}[htb]
\FigCap{Color-magnitude diagram of the upper part of red giant branch in
the LMC bar field. Thin solid lines indicate the magnitude of TRGB. Thick
solid line marks the boundary of the region used for determination of
TRGB magnitude.} 
\vskip5mm
\FigCap{Same as in Fig.~1 for the LMC NW field.} 
\vskip5mm
\FigCap{Same as in Fig.~1 for the SMC field.} 
\end{figure}
\end{document}